# Resolving the temporal dynamics of mode-locked laser with single-shot time-microscope

X. Liu [1, 2]

[1]*School of automation, Nanjing University of Information Science & Technology, Nanjing 210044, China*
[2]*College of Optical Science and Engineering, Zhejiang University, Hangzhou 310027, China*

**Abstract**

Mode-locked lasers, which produce ultrashort pulses in the picosecond and femtosecond range, have enabled some of the most precise measurements. However, despite significant recent progress, resolving the temporal behavior of their short pulses is still a challenge. State-of-the-art oscilloscopes with tens of picosecond resolution prevent time-resolved observations in mode-locked lasers and limit the real-time pulse evolution tracking of ultrafast lasers. Here, using the time-lens technique with a Raman amplifier, we implement an ultrafast single-shot time-microscope (TM) with a high temporal magnification factor of 355 and a time measurement window of 1 millisecond that contains $\sim 1.8 \times 10^4$ consecutive pulses. We use this TM to characterize the temporal evolution of mode-locked lasers and reveal a temporal sideway oscillation (winding) behavior, a previously unobserved feature of lasers in both theory and experiment. Our experimental observations confirm that the winding behavior is an essential feature in the operation of mode-locked lasers. We theoretically and experimentally found that the winding characteristic evolution originates from gain-induced fluctuations for relatively high gain energies, while Q-switched modulations being the main cause for lower energies. Our findings based on advanced real-time measurements open up new insights into ultrafast and transient optics and may impact future laser designs, modern ultrafast diagnostics, and influence progress in nonlinear optics in general.

**Introduction**

Passively mode-locked (PML) laser is an ideal platform to investigate complex dynamics, transient phenomena, nonlinear and ultrafast optical effects [1-5]. PML lasers can generate picosecond (ps) to femtosecond (fs) long optical pulses, widely used in optical communications, precise measurements, metrology, high-speed optical sampling, and data storage [3-12]. When operating in the steady-state regime, such lasers produce highly stable pulse trains with constant pulse energy (amplitude). This regime sometimes is called the continuous-wave (CW) mode-locking [1]. When deviated from the steady state, they can generate the amplitude-modulated pulse trains where the pulses are modulated periodically with a much longer Q-switched pulse envelope. This regime is known as Q-switched mode-locking [1,2,6]. This mechanism is in conjunction with Q-switching instabilities. Because of the gain saturation and recovery effect [13-16], variations of the pump power, and the spontaneous emission, PML lasers often operate in a regime characterized by pulse energy fluctuations. It is challenging to obtain absolutely stable mode-locking, i.e., constant pulse energy (amplitude), without implementing special stabilization technique. Therefore, PML laser operates in Q-switched or stable (CW) mode-locking regime according to the pumping strength (see Box 3 in Ref. [1]). Unfortunately, so far, no experiments have confirmed this theoretical prediction.

The buildup and evolution dynamics of PML lasers have been investigated experimentally and theoretically since 1990s [17-19]. More recently, their spectral dynamics and behavior have been measured using the time-stretch dispersive Fourier transform (DFT) technique [4, 20-22]. This technique opens wide range of opportunities for single-shot spectroscopy and real-time spectral measurements of ultrashort pulses. The use of this technique made it possible the experimental study of a number of ultrafast phenomena such as the internal motion of soliton molecules [4, 11], soliton explosions [23,24], soliton breathing [25], soliton pulsations [26], as well as rogue wave dynamics [27], modulation instability [28], dynamics of dissipative solitons [29], and the onset of PML lasers [30-34]. Unfortunately, the DFT technique only allows one to study the spectral dynamics [4]. Getting the temporal information in the time scales of ultrashort pulses remains highly challenging.

Commercial real-time oscilloscopes could only provide single-shot waveform measurement with resolution in the range of tens of ps. This is due to the microelectronic bandwidth limitations [35]. The temporal resolution can be significantly improved when taking advantage of all-optical techniques, for example, using the nonlinear process of four-wave mixing (FWM) [35]. In parallel with DFT-based spectral measurement technique, a time-lens technique (i.e., a temporal analogue of a spatial thin lens) has been proposed by using time-to-frequency conversion. This technique allows real-time pulse intensity measurements with sub-picosecond resolution [36, 37], e.g., a time-microscope (TM) has a temporal resolution of 250 fs [38]. In particular, with the use of silicon-chip-based ultrafast optical oscilloscope, a single-shot waveform measurement with 220 fs resolution has been achieved in Ref. [35]. Recently, Ryczkowski and Narhi *et al.* used a commercial time-lens magnifier with a temporal magnification factor of 76.4 and an effective temporal resolution of 400 fs over a time measurement window of 400 roundtrips. These achievements allowed them to measure the transient dissipative soliton dynamics, the spontaneous breather evolution, and the rogue wave events [29, 37], thus highlighting the characterization potential of this technique.

In this work, using time-lens and Raman amplifier combined with asynchronous technique, we implement an ultrafast single-shot TM with a magnification factor of 355, an effective temporal resolution of 76 fs, and a time measurement window of ~$1.8 \times 10^4$ roundtrips. Compared with previous reports [29, 37], the time measurement window and time amplification factor here are increased by about 50 times and 4.7 times, respectively. The Q-switched and stable (CW) mode-locking regimes of ultrafast laser operation are characterized and distinguished using the TM, the DFT, and the radio-frequency (RF) spectrum analyzer. The characterizations performed with our advanced equipment reveal a sideway oscillating (winding) behavior of PML lasers in time, a previously experimentally unobserved feature of lasers. Physically, the temporal winding characteristic originates from the gain-induced fluctuations (including gain saturation and recovery, pump power fluctuation, etc.) at higher pump power, but from the Q-switched modulation at lower pump power. Our results revealing a new distinctive feature of PML lasers will contribute to better understanding of laser dynamics. An important application of these results can be envisaged in improving stability of ultrafast lasers. We strongly believe that the advanced technique that allowed us to observe this phenomenon, i.e., simultaneous single-shot temporal and spectral characterization, will soon find a wider use as a standard measurement technique in ultrafast and transient optics.

**Results**

**Ultrafast single-shot TM**

The traditional time-lens technique usually avoids Raman amplification [39-45]. Here, we propose a TM implementation using a fiber-based time-lens and a Raman amplifier (Fig. 1a,

Supplementary Fig. S1, and Supplementary Note 1). To enhance the temporal magnification factor, the dispersion $D_i$ for the output propagation segment has to be increased so that the pulse amplitude is decreased. The Raman amplifier serves for compensating the reduction of the pulse amplitude. In such TM, the Raman amplifier combined with asynchronous technique can produce a measurement window much longer than in previous studies. Here, we achieve a single-shot acquisition TM with a highly temporal magnification factor of 355, an effective temporal resolution of 76 fs and a record length of ~400 ps over a time measurement window of ~$1.8\times10^4$ roundtrips (i.e., 1 ms time interval). The detailed description of the TM is given in the Supplementary Note 4. Our TM has much longer time measurement window than that reported in Refs. [29][37], where the time-lens had a time measurement window of only 400 roundtrips (20 μs time interval). With our advanced apparatus, we can measure the dynamic pulse evolution of a PML laser on a much larger time-scale, which increased from 20 μs in previous studies to 1 ms in our case (~50 times). These significant technical improvements allow us to discover experimentally the temporal winding behavior in mode-locked lasers, a so far unobserved feature both theoretically and experimentally.

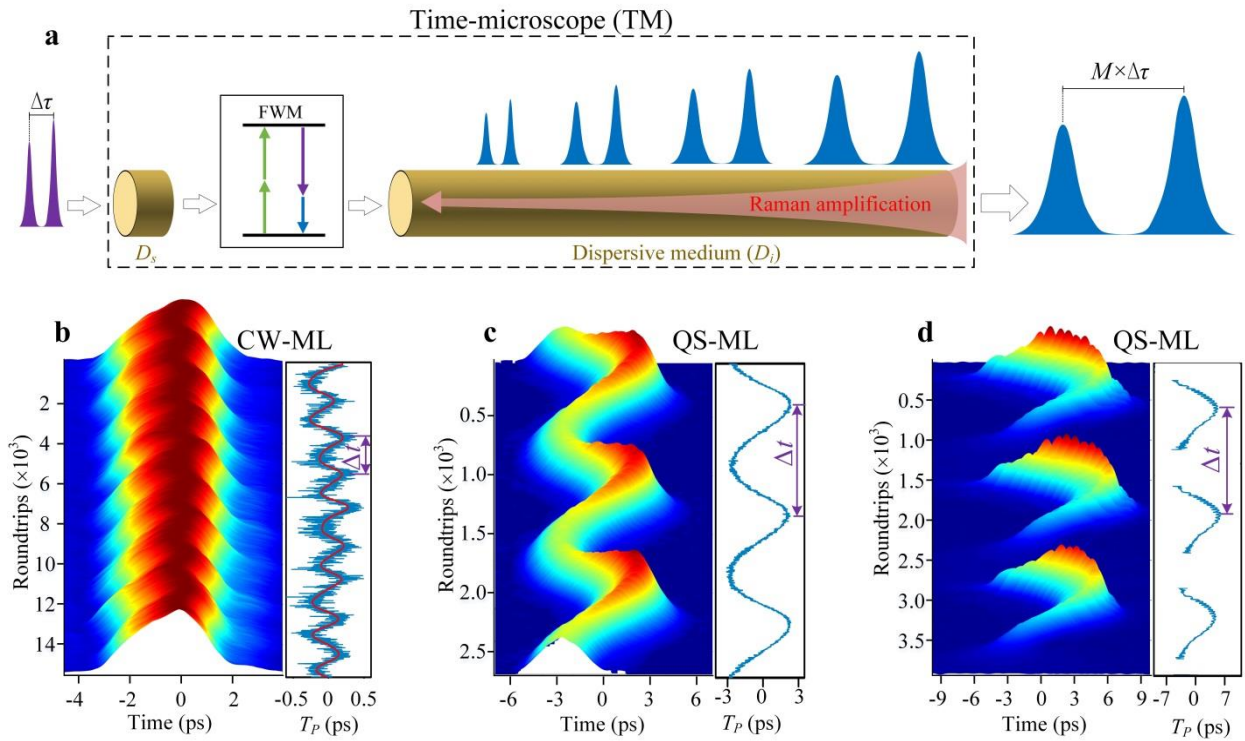

**Fig. 1 Ultrafast single-shot temporal TM and its application to the real-time measurements of dynamic pulse evolution in a PML laser. a** Schematic diagram of the single-shot temporal TM. It is made of a FWM-based parametric time-lens and a Raman amplifier, featuring the temporal magnification factor $M$ of 355, the temporal resolution of 76 fs, and the record length of ~400 ps. The Raman amplification and the asynchronous technique contribute to the achievement for such TM with time measurement window of ~$1.8\times10^4$ roundtrips (i.e., 1 ms). **b-d** Experimental real-time single-shot TM measurements for stable (CW) and Q-switched mode-lockings at pump power $P$=50.8, 47.9, and 46.9 mW (from left to right). Our experiments show a winding temporal evolution of the PML pulses in contrast to the straight roundtrip propagation understood so far. In each panel, the plots show: the measured temporal intensity profile on the left, and the extracted temporal location ($T_p$) of the peak power on the right. A red curve in **b** is a fitting curve. Each plot contains (**b**) ~$15.5\times10^3$, (**c**) ~$2.7\times10^3$, and (**d**) ~$3.9\times10^3$ roundtrips, respectively. Roundtrips mean that the recorded time series is segmented with respect to the roundtrip time $T_{rep}$ of 54.8107 ns. In this representation, the pulses of the PML laser are winding with a modulation period of (**b**) 1772 roundtrips (i.e., $\Delta t$~97.1 μs), (**c**) 945 roundtrips (i.e., $\Delta t$~51.8 μs), and (**d**) 1284 roundtrips (i.e., $\Delta t$~70.4 μs). The amplitude of this back and forth winding motion is (**b**) ~0.25 ps, (**c**) ~2.8 ps, and (**d**) ~6 ps. CW-ML, stable continuous-wave (CW) mode-locking; QS-ML, Q-switched mode-locking.

### Temporal winding evolution of PML lasers

In experiments, the TM and the DFT are used to temporally and spectrally resolve the evolution dynamics of a PML laser, respectively. The laser generates pulses at the central wavelength of ~1551 nm with the duration of ~3 ps and the repetition rate of 18.2446 MHz (corresponding to the cavity roundtrip time $T_{rep}$ of 54.8107 ns). The pulsed laser output is recorded using a high-speed photodetector and a real-time oscilloscope. Single-shot data are simultaneously measured via direct measurement, TM, and DFT (Supplementary Fig. S1), corresponding to direct, temporal, and spectral intensity profiles, respectively. Each data set is recorded over a duration of 1 ms and contains about $1.8\times10^4$ consecutive pulses (one pulse per $T_{rep}$). The recorded time series are segmented with the length of each segment $T_{rep}$, yielding a 2×2 matrix representation with the x-axis corresponding to information within a single roundtrip and the y-axis depicting the dynamics across consecutive roundtrips [33]. Three types of measurements are as follows. Temporal dynamics of PML laser are obtained using TM (Fig. 1b-d), direct detection of the output train (without TM and DFT) generates direct timing information (Fig. 2a-c), and spectral profiles are achieved by a time–wavelength mapping (DFT) via inserting a dispersion-compensating fiber (Fig. 2d-f).

The experimental results show that the laser operates in unstable regime at the pump power $P$<46.9 mW, Q-switched mode-locking at $P$=46.9~48.5 mW, stable (CW) mode-locking at $P$=48.7~55.3 mW, and unstable or multi-pulse state at $P$>55.3 mW (Fig. 3d and Supplementary Fig. S15), respectively. Figure 1b-d shows the experimental real-time measurements via our single-shot TM for stable (CW) and Q-switched mode-lockings at $P$=50.8, 47.9, and 46.9 mW, respectively. Here, each plot contains ~15.5×10$^3$ (Fig. 1b), ~2.7×10$^3$ (Fig. 1c), and ~3.9×10$^3$ (Fig. 1d) roundtrips. Note that Fig. 1d and Fig. 2c,f illustrate Q-switched mode-locking at $P$=46.9 mW, being taken from a data sequence just after starting such mode-locking regime.

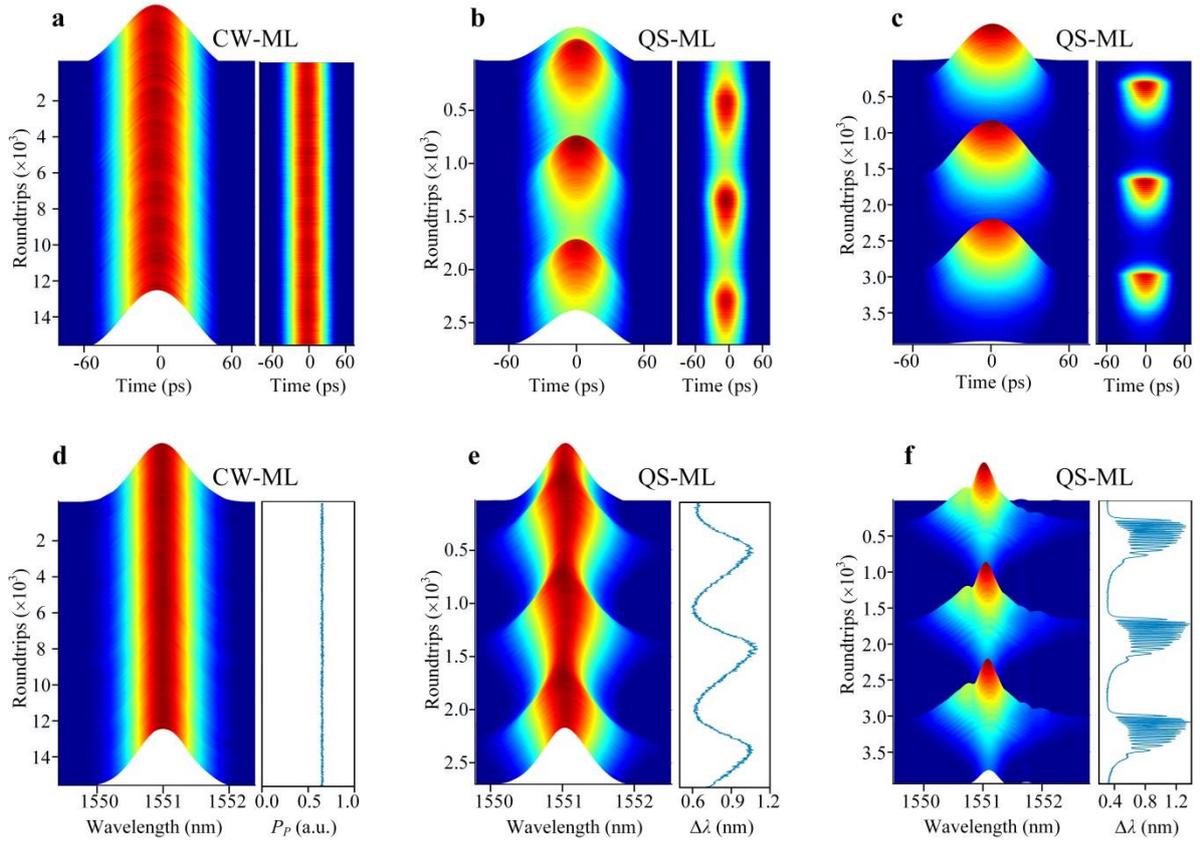

**Fig. 2 Experimental real-time characterizations of a PML laser, using direct measurement and DFT, for Q-switched and stable (CW) mode-locking regimes. a-c** Experimental real-time measurements recorded directly by a real-time oscilloscope (without using TM and DFT) at pump power $P$=50.8, 47.9, and 46.9 mW, respectively. In each panel, the plots show: the measured temporal intensity profile on the left, and the top-view graph on the right. **d-f** Experimental real-time measurements with the DFT. In each panel, the plots show: the measured spectral intensity profile on the left, and (**d**) the peak power ($P_p$) and (**e** and **f**) the spectral 3-dB bandwidth ($\Delta\lambda$) on the right. When the laser is in the stable (CW) mode-locked state, the pulse profiles at $P$=50.8 mW are simultaneously acquired by TM (Fig. 1b), direct measurement (**a**), and DFT (**d**), respectively. Here, the panel (**a**) shows the temporal while the panel (**d**) shows the spectral intensity profiles (including the peak power and the bandwidth). Both are uniform along the roundtrip number when the direct measurements and the DFT are used. The TM technique actually reveals the pulse winding (Fig. 1b). When operating in the Q-switched mode-locked state at $P$=47.9 and 46.9 mW, the temporal (**b** and **c**) and the spectral (**e** and **f**) intensity profiles evolve periodically. However, the direct measurement and the DFT techniques do not have sufficient resolution to reveal the winding behavior of the mode-locked laser pulses, and then the location of the peak power along the roundtrip is at a fixed position (straight propagation). On the contrary, the TM technique reveals the winding behavior of the actual temporal evolution (Fig. 1c,d).

Because of the gain-induced fluctuation and the influence of Q-switched modulation, the laser peak power and its central wavelength can vary from one roundtrip to another. Then the roundtrip time varies itself with an associated changeable temporal location ($T_p$) of pulse peak. The pulses shifting their temporal position periodically can be called swinging. When such pulses in successive roundtrips are plotted one behind the other, this dynamics can be seen as `winding'. Our high-resolution experiments clearly demonstrate that the temporal evolution of pulses in the PML laser is indeed winding rather than having a fixed position along the roundtrip. Three typical sequences of temporal intensity profiles are shown in Fig. 1b-d. Their RF spectra and oscilloscope traces are shown in Fig. 3a-c and Supplementary Fig. S14a,d,e, respectively. The PML pulses for the above three examples have the winding period of 1772, 945, and 1284 roundtrips, corresponding to the modulation period $\Delta t$ of about 97.1, 51.8, and 70.4 µs (Fig. 3d) and the modulating frequency $\Delta f$ (i.e., frequency difference) of 10.3, 19.3, and 14.2 kHz (Fig. 3a-c). The corresponding swing of the pulse peak amplitude in time is about 0.25, 2.8, and 6 ps, respectively. These oscillations are shown on the right-hand side panels of Fig. 1b-d.

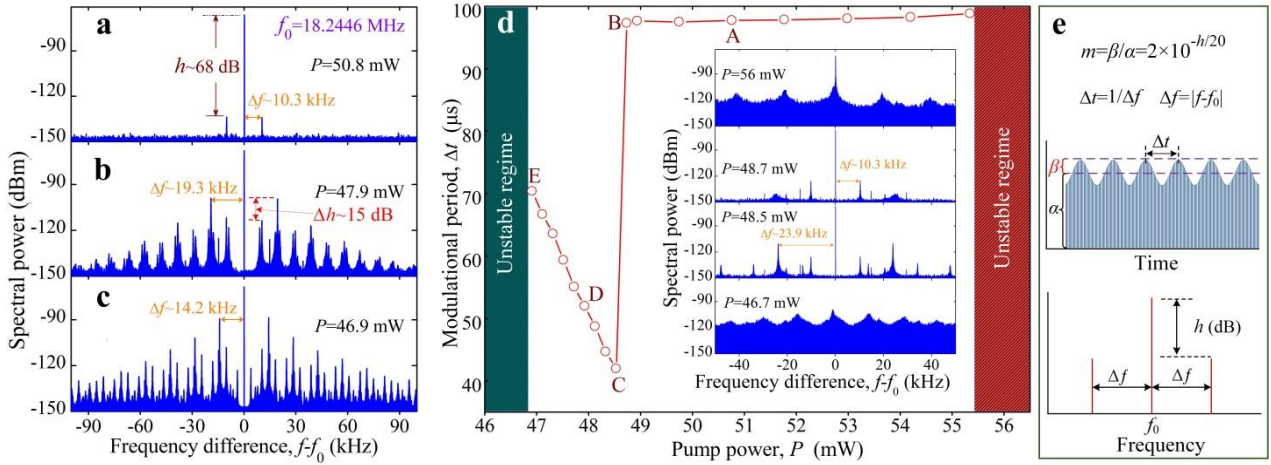

**Fig. 3 Radio-frequency (RF) spectra and modulation period of a PML laser. a-c** Fundamental RF spectra of laser with 200 kHz span at pump power $P$=50.8, 47.9, and 46.9 mW (from top to bottom). The fundamental frequency $f_0$=18.2446 MHz corresponds to the round trip time $T_{rep}$=54.8107 ns, i.e., $f_0$=1/$T_{rep}$. The corresponding real-time oscilloscope traces are shown in Supplementary Fig. S14a,d,e. **d** Relationship between the modulation period $\Delta t$ and the pump power $P$. Inset: RF spectra at $P$=56, 48.7, 48.5, and 46.7 mW. The laser operates in the unstable regimes for $P$>55.4 mW and $P$<46.9 mW. $\Delta t$ is the reciprocal of the frequency difference $\Delta f$, i.e., $\Delta t$=1/$\Delta f$. The points corresponding to $\Delta t$ at $P$=50.8, 48.7, 48.5, 47.9, and 46.9 mW are marked as A, B, C, D, and E, respectively. **e** (Top) The pulse sequence with the modulation period $\Delta t$, the carrier amplitude $\alpha$, and the modulating amplitude $\beta$. (Below) The corresponding RF spectrum with the modulating frequency $\Delta f$ (i.e., frequency difference $\Delta f$=|$f$-$f_0$|).

**Spectral analysis for the time series of laser pulses**

Laser pulses (via a photodetector) can be displayed either in the time domain, using an oscilloscope, or in the frequency domain using a RF spectrum analyzer. Typically, oscilloscopes provide higher resolution in time but lower amplitude resolution. The latter can be offset by RF spectrum analyzers because they provide higher amplitude resolution [46-48]. Generally, the RF spectrum analyzer enables accurate measurement of the modulation period, amplitude, and frequency. When the modulating signal evolves with a cosine function, the modulation period $\Delta t$ and the modulating frequency $\Delta f$ are inversely proportional to each other,

$$\Delta t = 1/\Delta f, \tag{1}$$

while the modulation factor, $m = \beta/\alpha$, is expressed as follows

$$m = 2 \times 10^{-h/20}. \tag{2}$$

Here, $\alpha$ and $\beta$ are the amplitudes of the carrier and the modulating signals (Fig. 3e and Supplementary Fig. S6a), respectively. $h$ is the difference between the intensities of the spectral components at the fundamental frequency and the sidebands (Fig. 3a,e and Supplementary Fig. S6b).

Generally, the amplitude modulation of laser pulses originates from many factors, which cause deviation from a typical cosine waveform. These factors (e.g., gain-induced fluctuation, Q-switched modulation, and other perturbations) can be accurately distinguished using the RF spectra. They are marked with multiple frequency lines at different positions in Supplementary Fig. S7b and Supplementary Fig. S13d, showing the schematic diagram and the experimental results, respectively. The amplitude ratio between the subordinate and dominant modulating signals is given by

$$R_{amp} = 10^{-\Delta h/20}, \tag{3}$$

where $\Delta h$ is the relative intensity difference between two modulating signals in the RF spectrum (Fig. 3b and Supplementary Fig. S7b). The detailed derivation for Eq. 1 to 3 is shown in the Supplementary Note 5.

**Comparison of results recorded by direct measurements, DFT, and TM**

When the PML laser is in the stable (CW) mode-locking regime, both the direct measurement and the DFT show that the temporal (Fig. 2a) and spectral (Fig. 2d) intensity profiles are uniform along the roundtrip. Their peak power and bandwidth are constant with respect to the roundtrip. Therefore, the fact that the temporal evolution of the PML laser is winding cannot be seen from the direct measurement and from the DFT. In contrast, our TM can reveal the true characteristics of temporal dynamics in a PML laser, which exhibits the winding behavior (Fig. 1b) rather than the straight propagation along the roundtrip. The pulse peak swings with the period $\Delta t$ of ~97.1 μs (i.e., 1772 roundtrips) and the oscillating amplitude of ~0.25 ps (see the right panel in Fig. 1b). To validate this phenomenon, the RF spectra are measured, as shown in Fig. 3a. This plot shows the frequency difference (i.e., modulating frequency) $\Delta f$ to be ~10.3 kHz that is the reciprocal of $\Delta t$=97.1 μs (according to Eq. 1). The difference between the amplitudes of the fundamental frequency and the sideband, $h$, is ~68 dB. Thus, the modulation factor, $m$, is about $8 \times 10^{-4}$ according to Eq. 2. This shows that the amplitude fluctuations of laser pulses in the stable (CW) mode-locking state are very small.

When the laser is in the Q-switched mode-locking regime, the temporal (Fig. 2b,c) and spectral (Fig. 2e,f) intensity profiles are periodic functions of the roundtrip number if only DFT or the direct measurement is used. Such phenomenon is known as breathing soliton [25] or soliton pulsations [26], where the winding characteristic is not revealed. Just as in the stable (CW) mode-locked regime, the TM unveils that the temporal pulse dynamics at the Q-switched mode-locking are winding (Fig. 1c,d), rather than having a fixed temporal position, along the roundtrip. The pulses swing to right and left with $\Delta t$~51.8 μs (i.e., 945 roundtrips) and ~70.4 μs (i.e., 1284 roundtrips) and with the amplitude of ~2.8 and ~6 ps, respectively, as shown at the right-hand side of Fig. 1c,d. The corresponding RF spectra (Fig. 3b,c) show that $\Delta f$ are ~19.3 and ~14.2 kHz, respectively, which are the reciprocal of $\Delta t$=51.8 and 70.4 μs. Therefore, the RF spectra recorded by the spectrum analyzer confirm the results recorded by our TM.

**Theoretical confirmation**

When a laser is pumped, ions are excited from the ground state to the excited state and finally return to the ground state by emitting photons. The transient features (e.g., relaxation oscillation) of this process can be accurately described by the rate equations [49] (see Supplementary Note 6). An intra-cavity saturable absorber could generally initiate the passively Q-switched operation that can be predicted by the coupled rate equations (see Supplementary Note 7). Numerical calculations show that, when the time $t$ is large enough (e.g., $t$>400 μs), the modulation factor approaches infinitesimal and the photon number in the laser cavity is close to a constant without fluctuation at a fixed pumping rate (Supplementary Fig. S8c). In practice, the pumping rate usually varies with time due to the gain saturation and recovery [13-16] and the fluctuation of the pump power. The experimental results illustrate that the gain saturation and recovery time can be about 100 μs (from dozens to hundreds of μs). Therefore, the gain-induced fluctuations (including gain saturation and recovery, pump power fluctuations, etc.) can cause small changes in the pulse energy and its peak power (see Supplementary Fig. S7 and Supplementary Fig. S13d: the schematic diagram and the experimental results).

Here, we use the amplitude modulation model with a cosine function to describe the modulation of the pumping rate $W(t)$, as shown in Fig. 4a and Supplementary Fig. S9a. The pumping rate evolves with a modulation factor $m_w$ (i.e., relative fluctuation) of 7.97×10$^{-4}$ and a modulation period of 97.1 μs. The numerical results show that the photon number varies with the same envelope and period of the pumping rate (Fig. 4b or Supplementary Fig. S9b), i.e., the fluctuation of photon number follow the cosine function with the modulation period $\Delta t$ of 97.1 μs. The modulation factor of the pumping rate is approximately equal to that of the photon number, i.e., $m_w \approx m_p$ (Supplementary Fig. S9d). The spectral information of Fig. 4b in the frequency domain can be achieved by Fourier transformation (Supplementary Note 5), as shown in Supplementary Fig. S9c (here, $\Delta f$ is ~10.3 kHz and $h$ is ~68 dB). Theoretical results (Supplementary Fig. S9c) are consistent with experimental observations (Fig. 3a). Numerical results show that $\Delta t$ for the relaxation oscillation remains almost constant with respect to the pumping rate $W$ (Fig. 4c or Supplementary Fig. S9e) while $\Delta t$ for the Q-switched operation decreases almost linearly to the volumetric pumping rate $R_P$ (Fig. 4d or Supplementary Fig. S10c). These results are in good agreement with the experiments (Fig. 3d or Supplementary Fig. S15). The detailed calculations are given in Supplementary Notes 6 and 7.

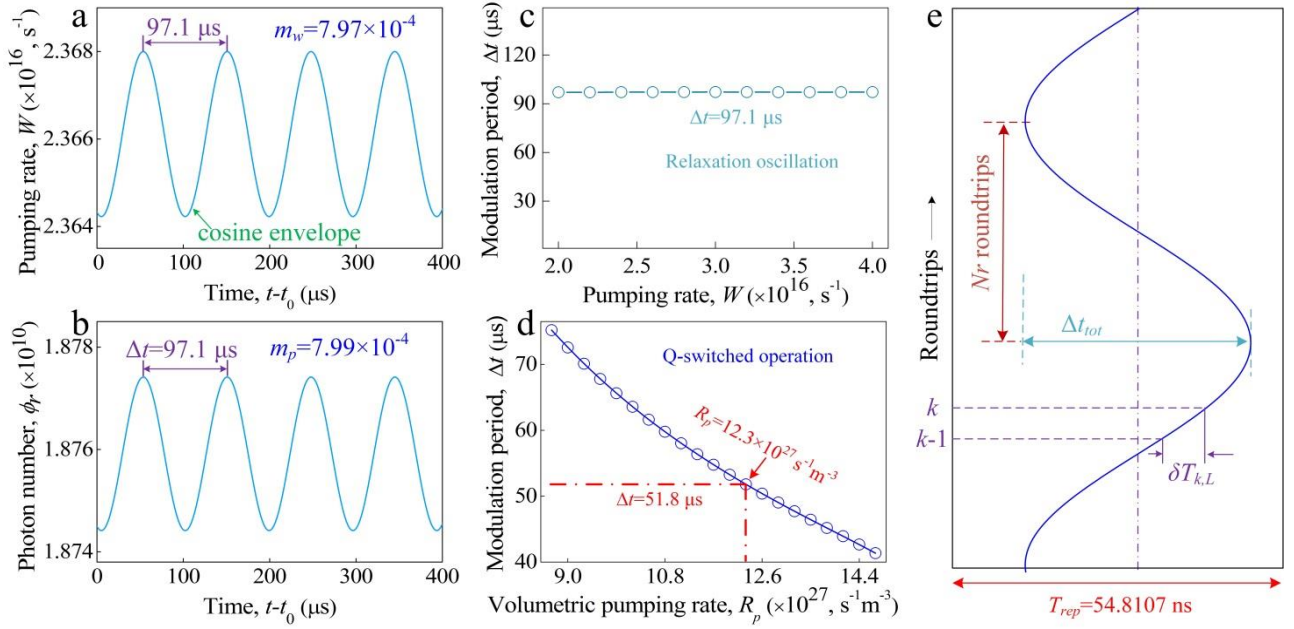

**Fig. 4 Theoretical results. a** Modulation of the pumping rate $W(t)$ with cosine function, relative fluctuation $m_w$ of $7.97 \times 10^{-4}$, and modulation period of 97.1 μs. $t_0$ is the reference time. **b** Fluctuation of photon number, achieved by simulating Eqs. 5 and 6 with the modulation of pumping rate $W(t)$ in **a**. The fluctuations of the photon number evolve as cosine function with the relative fluctuations $m_p = 7.99 \times 10^{-4}$ and the modulation period $\Delta t = 97.1$ μs. **c** Relationship between $\Delta t$ and $W$ for the relaxation oscillation. **d** Relationship between $\Delta t$ and $R_P$ (volumetric pumping rate: $R_p = W/V_P$; see Supplementary Note 7 for further details) for the Q-switched operation of laser. When simulating curves in **c** and **d**, $m_w$ remains unchanged while $W$ and $R_P$ are variable. $\Delta t$ remains nearly a constant ($\Delta t \sim 97.1$ μs) when $W$ is changed in the relaxation oscillation case (**c**), while it decreases roughly linearly with $R_P$ in the Q-switched operation regime (**d**). When $R_p = 12.3 \times 10^{27}$ s$^{-1}$m$^{-3}$, $\Delta t$ is ~51.8 μs that is in agreement with the experimental result (Fig. 3d). **e** Total time difference, $\Delta t_{tot}$, after $N_r$ roundtrips. $\delta T_{k,L}$ is the time difference between the two adjacent roundtrips (from $k$-1-th to $k$-th). The total length of the horizontal axis is equal to the reference roundtrip time $T_{rep} = 54.8107$ ns.

Due to the fluctuations of the central wavelength and the peak power of laser pulses, the time difference between two adjacent roundtrips, $\delta T_{k,L}$, is variable rather than zero (see Methods, Supplementary Note 8, Fig. 4e, and Supplementary Fig. S12). After $N_r$ roundtrips, the total time difference $\Delta t_{tot}$ is given by

$$\Delta t_{tot} = \sum_{i=1}^{N_r}\left[\delta n_i + \frac{\Delta P_i \gamma \lambda_i}{2\pi}\right]\frac{L}{c}, \qquad (4)$$

where $\gamma$ is the nonlinear coefficient, $L$ is the length of laser cavity, $c$ is the speed of light in vacuum, and $\lambda_i$ is the central wavelength of the pulse at the $i$-th roundtrip. $\delta n_i$ and $\Delta P_i$ are the differences of the refractive index and power between the $i$-th roundtrip and the reference roundtrip, respectively. The detailed derivation can be found in Supplementary Note 8. Numerical results show that $\Delta t_{tot}$ is ~0.5 ps at the wavelength difference $\delta\lambda \approx 1.2\times10^{-3}$ nm, $\Delta P=0$, and $N_r=886$ (Supplementary Fig. S12a), as well as ~5.6 ps at $\Delta P \approx 1000$ W, $\delta\lambda=0$, and $N_r=473$ (Supplementary Fig. S12b). Note that the DFT cannot reveal the wavelength fluctuations as low as $\delta\lambda \approx 1.2\times10^{-3}$ nm (Fig. 2d).

**Discussion**

To further understand the details of the laser pulse evolution, a part of Fig. 1d and Fig. 2c,f is redrawn as Supplementary Fig. S16. Obviously, the direct measurement and the DFT can only reveal rough phenomena like pulse breathing (Supplementary Fig. S16b,c). On the contrary, our TM can unveil the winding dynamics (Supplementary Fig. S16a), in addition to the pulse breathing. We further truncate an interval of 200 roundtrips from Fig. 1d and Fig. 2f, which are redrawn as Fig. 5a,b. Here, our experimental results are similar to the previous observations (Fig. 3e,f in Ref. [29]). Namely, the results clearly demonstrate the pulse breathing. However, it is difficult to see the winding dynamics of temporal evolution due to relatively smaller number of roundtrips (here, 200 roundtrips). The higher resolution of our equipment (here, 76 fs temporal resolution and 0.14 nm spectral resolution) allowed us to observe significant modulation in both the measured spectral and temporal amplitudes. Our results can even reveal the pulse amplitude growth and unveil the aperiodic modulation along the roundtrip number axis. As we can see, the experiments here are more detailed and provide better insight into the pulse dynamics than the previous observations [29].

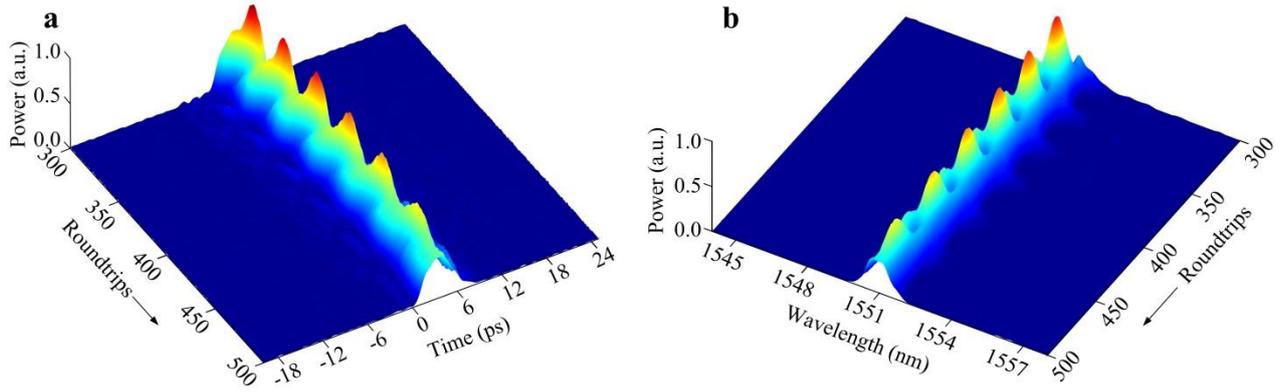

**Fig. 5 Real-time spectral and temporal characterization of Q-switched mode-locking regime near stability limit**. Experimental measurements over 200 roundtrips for (**a**) temporal and (**b**) spectral intensity profiles at *P*=46.9 mW. These plots are truncated from Fig. 1d and Fig. 2f, respectively. The Q-switched modulation and the breathing behavior are clearly seen, but the winding motion of temporal profile is too small to be distinguished in **a**. This is in agreement with the results of previous observations (Fig. 3e,f in Ref. [29]). In fact, the temporal evolution appears winding rather than straight over a larger number of roundtrips (e.g., 900 roundtrips) (see Supplementary Fig. S16a for details).

Figure 3d and Supplementary Fig. S15 illustrate that the PML laser produces Q-switched and stable (CW) mode-lockings at the pump power *P* of ~46.9 to 48.5 mW and ~48.7 to 55.3 mW, respectively. The Q-switched modulation has an important influence on the Q-switched mode-locking regime where $\Delta f$ is approximately linear along $R_p$ or *P* (Supplementary Fig. S10b and Supplementary Fig. S15; theory and experiment). However, $\Delta f$ is approximately a constant along *W* or *P* at stable (CW) mode-locking (Supplementary Fig. S9e and Supplementary Fig. S15; theory and experiment). Obviously, our experiments are in good agreement with the previous results (see Box 3 in Ref. [1]), i.e., the PML laser operates in Q-switched or stable (CW) mode-locking regimes according to the gain saturation (corresponding to *P*).

Supplementary Fig. S13 shows that the frequency component at $\Delta f \approx 10.3$ kHz always exists in both Q-switched and stable (CW) mode-locking regimes. Our experimental observations suggest that the gain-induced fluctuation (e.g., gain saturation and recovery effect and/or fluctuation of pump power) is an inherent characteristic of mode-locked lasers. In fact, the relaxation oscillation is an intrinsic physical process of mode-locked lasers [30-32]. The theoretical results confirm that the relaxation oscillation under the condition of gain-induced fluctuation contributes to the power fluctuation of laser pulses, which is independent of the pumping rate (Fig. 4c and Supplementary Fig. S9e).

The RF spectra are measured by a RF spectrum analyzer (R&SFSW26) together with a high-speed photodetector. The displayed average noise level (DANL) of such analyzer can be as low as -169 dBm [46], which is much lower than the noise level of the laser input (about -150 dBm). The low DANL is critical to reveal weak sideband modulating signals of stable CW mode-locking, whose value is about -135 dBm at $\Delta f$~10.3 kHz (Fig. 3a and Supplementary Fig. S17a). For comparison, we have measured the same laser output using another RF spectrum analyzer (Keysight E4440A), whereby the weak sideband is covered under relatively high DANL (Supplementary Fig. S17b).

## Materials and methods
### Experimental setup

The schematic diagram of our experimental setup is shown in Supplementary Fig. S1. It consists of a PML laser, a tunable laser pump, a TM (including the time-lens and Raman amplifier sections), and the measuring equipment. The pulses generated by the PML laser are simultaneously recorded by direct measurement, DFT, and TM. The DFT provides the ability of real-time single-shot spectral measurement with the spectral resolution of ~0.14 nm. The TM can magnify the signal by a factor of $M$=355 and provide the ability of real-time single-shot temporal waveform measurement with the temporal resolution of ~76 fs. The record length of TM is determined by the duration of the pump pulse, which is ~400 ps. The Raman amplification can compensate the attenuation of idler, which is used to increase the time measurement window up to 1 ms (~$1.8 \times 10^4$ roundtrips).

The PML fiber laser is comprised of a carbon nanotube saturable absorber (CNT-SA), a segment of erbium-doped fiber (EDF), a polarization controller (PC), a bandpass filter (BPF2), some single-mode fiber (SMF) pigtails, and a polarization-independent hybrid combiner of a wavelength division multiplexer, a tap coupler, and an isolator (WTI). The gain medium is ~1 m-long EDF with 110 dBm peak absorption at 980 nm, pumped by a laser diode (LD). The total length of the laser cavity is ~11.3 m, corresponding to the fundamental repetition rate $f_0$ of 18.2446 MHz and the cavity roundtrip time $T_{rep}$ of 54.8107 ns. The DFT uses 5-km dispersion compensating fiber (DCF, $D=$ −160 ps/(nm km)) that provides the ability of real-time single-shot spectral measurement with the resolution of ~0.14 nm[31]. The time-lens uses two coils of DCF (~100 m for $D_s$ and ~35.5 km for $D_i$) and ~100 m highly nonlinear fiber (HNLF, near-zero dispersion at 1550 nm, nonlinear coefficient $\gamma$ of ~10 $W^{-1}km^{-1}$). The pump seed for the time-lens is generated from a CNT-based mode-locked fiber laser with tunable repetition rate. The seed is amplified through a self-similar amplification scheme, filtered by the BPF1 (central wavelength: 1560 nm, bandwidth: 10 nm), and stretched by ~200 m DCF ($D_p$). The pump duration is ~400 ps after stretching by the DCF and the pump pulse can impose a quadratic temporal phase through FWM. The idler generated through FWM is filtered by a tunable bandpass filter (TBPF) and amplified by a Raman amplifier.

**Raman amplification and its application on time-lens**

Without the Raman amplification, the time measurement window and the temporal magnification factor $M$ of the time-lens are limited by the nonlinear conversion efficiency, the attenuation of idler, and the pulse broadening, as shown in Supplementary Fig. S2. This would limit the time-lens measurement windows to several hundreds of roundtrips [29, 35, 41, 44], as well as the DCF length to less than 17 km (see details in Supplementary Note 2).

Raman amplification can compensate the idler attenuation and increase its peak power for longer propagation distance, as shown in Supplementary Fig. S3b. TM with Raman amplification possesses larger time measurement window and provides the capability of higher $M$. Numerical results show that, under the condition of the same conversion efficiency and peak power, the time magnification factor $M$ increases ~9.5 times (Supplementary Fig. S3c-1,c-3). Therefore, the Raman amplification contributes to much larger value of $M$ in the case of our TM (see details in Supplementary Note 3).

**Measurement method**

A RF spectrum analyzer (R&SFSW26) with the displayed average noise level (DANL) as low as -169 dBm and a high-speed real-time oscilloscope (R&S®RTP164) with the 40

Gsamples/s sampling rate and the bandwidth of 16 GHz are used to measure the laser output performances.

**Relaxation oscillation with the unchanged modulation period $\Delta t$**

The instantaneous state of fiber laser is determined by the populations in the upper level of the signal transition and the photons in the resonator [50]. The transient features can be expressed by the rate equations[49, 50], i.e.,

$$\frac{dn_r}{dt} = W - n_r B_s \phi_r - \frac{n_r}{\tau_u}, \tag{5}$$

$$\frac{d\phi_r}{dt} = n_r B_s \phi_r - \frac{\phi_r}{t_c} + \frac{n_r}{p_m \tau_u}. \tag{6}$$

Here, $n_r$ and $\phi_r$ represent the population inversion in the gain medium and the photon number in the laser cavity, respectively. $\tau_u$ is the lifetime of the upper-laser level, given by $\tau_u = \tau_1 + \tau_2$ (Supplementary Fig. S8a). $W$ is the pumping rate. $t_c$ is the photon lifetime, related to the cavity loss. $B_s$ is the Einstein coefficient, given by $B_s = 1/(p_m \tau_u)$ [50]. $p_m$ is the number of cavity modes coupled to the fluorescent line. The numerical results show that the modulation period of photon number fluctuation, $\Delta t$, is almost independent of the pumping rate $W$ (Fig. 4c and Supplementary Fig. S9e) (see details in Supplementary Note 6).

**Q-switched modulation with the linearly changed modulation period $\Delta t$**

The passively Q-switched laser system can be considered as a laser medium with an intra-cavity SA. The coupled rate equations can be written as [51-53]

$$\frac{d\phi_d}{dt} = \frac{\phi_d}{T_{rep}} \left[ \sigma n_d L_{EDF} - \sigma_{gs} n_{gs} L_s - \sigma_{es}(n_{0s} - n_{gs})L_s - \delta \right], \tag{7}$$

$$\frac{dn_d}{dt} = R_p(1 - \frac{n_d}{N_T}) - 2\sigma c \phi_d n_d - \frac{n_d}{\tau_u}, \tag{8}$$

$$\frac{dn_{gs}}{dt} = \frac{n_{0s} - n_{gs}}{\tau_{gs}} - \sigma_{gs} c \phi_d n_{gs}. \tag{9}$$

Here, $\phi_d$, $n_d$, and $n_{gs}$ denote the photon density, the instantaneous population inversion density, and the instantaneous population density of the absorbing state of SA, respectively. $\sigma$ is the laser-stimulated emission cross-section. $\sigma_{gs}$ and $\sigma_{es}$ are the ground- and excited-state absorption cross-sections of absorber, respectively. $L_{EDF}$ and $l_s$ are the lengths of gain and absorbing media, respectively. $\delta$ is the cavity loss. $N_T$ is the doping concentration of gain medium. $n_{0s}$ is the total population density of SA. $\tau_{gs}$ is the spontaneous emission lifetime of the absorber. $R_p$ is the volumetric pumping rate, given by $R_p = W/V_p$ ($V_p$ is the effective mode volume of laser). The numerical results show that the modulating frequency $\Delta f$ (i.e., the frequency difference of the photon density) is approximately proportional to $R_P$ (Supplementary Fig. S10b) while the modulation period of photon density, $\Delta t$, is approximately inversely proportional to $R_P$ (Fig. 4d and Supplementary Fig. S10c) (see details in Supplementary Note 7)

**Roundtrip time fluctuation induced by the refractive index**

In the fiber, the roundtrip time $T_L$ of laser is given by [54]

$$T_L = [n(\lambda) + \Delta n] L/c . \tag{10}$$

Here, $n(\lambda)$ is the linear part of the refractive index, given by the Sellmeier equation [55]. $\Delta n$ is the nonlinear refraction, referring to the intensity dependence of the refractive index. We define the reference roundtrip time as $T_{rep}$ (here $T_{rep}$=54.8107 ns) at the peak power of $P_0$ and the central wavelength of $\lambda_0$. The time difference between two adjacent roundtrips (from $k$-1-th to $k$-th roundtrips) is given by

$$\delta T_{k,L} = \left[ \delta n_k + \frac{\Delta P_k \gamma \lambda_k}{2\pi} \right] \frac{L}{c} . \tag{11}$$

Here, $\Delta P_k = P_k - P_0$, $\delta n_k = n(\lambda_0 + \delta\lambda_k) - n(\lambda_0)$, and $\delta\lambda_k = \lambda_k - \lambda_0$. The schematic diagram of Eq. 11 is shown in Fig. 4e. The detailed derivation is shown in Supplementary Note 8.

**Authors**: X. Liu, D. Popa, N. Akhmediev, J. Zheng, *et al.*,

**Acknowledgments.** We thank S. Liu and R. Su for fruitful draft and revision in Supplement. We thank W. Geng for fruitful discussions and C. Jin for helpful technical assistance.